\documentclass[12pt]{article}
\usepackage{epsfig}
\usepackage{graphicx}
\begin{document}
\setlength{\baselineskip}{.6cm} \setlength{\parskip}{1mm}

\begin{titlepage}
\bigskip
\begin{flushright}
   {13.03.2004\\
   WIC/07/04-MAR-DPP}
\end{flushright}

\bigskip \bigskip
\begin{center}{\LARGE\bf A New Algorithm for Inclusive Search of SUSY Signals.}
\end{center}
\bigskip

\begin{center}{Ehud Duchovni \footnote{ehud.duchovni@weizmann.ac.il}, Eugene Prosso and Peter
Renkel \footnote{Renkel@wisemail.weizmann.ac.il} }\end{center}

\begin{center}{\it Particle Physics Department,\\
Weizmann Institute of Science, Rehovot 76100, Israel}
\end{center}

\bigskip\bigskip\bigskip

\begin{abstract}
A new algorithm designed to reduce the model dependence in future
SUSY searches at the LHC is described. This algorithm can
dynamically adapt itself to a wide range of possible SUSY final
states thus reducing the need for detailed model-driven analysis.
Preliminary study of its performance on simulated MSSM, GMSB and
AMSB final states is described, and a comparison with traditional
search procedures, whenever available, is performed.

\end{abstract}
\end{titlepage}
%
%
\newcommand{\qq}         {\mbox{$Q^2$}}

\section{Introduction}
While the case for nature to be supersymmetric is very appealing,
the understanding of the way in which Supersymmetry is broken is
far from being established. The details of this symmetry breaking
determine the SUSY mass spectrum and consequently the way in which
SUSY will exhibit its existence at the LHC. An attempt to perform
a virtually model independent search for a large class of possible
SUSY final states is reported in this note. The outlines of the
proposed wide-scope algorithm are presented in the next section.
The widening of the scope of the search is achieved by dynamic
adaptation of the algorithm to the peculiarities of the signal.
Such a procedure is likely to result in a reduction of the search
sensitivity when compared to sophisticated dedicated analysis
techniques like Artificial Neural Networks (ANN), which are based
on a prior knowledge of the signal characteristics but the
deterioration is shown to be marginal and the algorithm performs
significantly better than simple cuts. It is argued that a
combination of the traditional Model driven searches and the
present wide scope procedure will allow ATLAS to conduct the most
effective search for SUSY (and probably other) final states. These
statements are substantiated by the MC studies that are described
in the following sections.

\section{Description of the technique}
The exact nature of the expected SUSY final states depends on the
details of the way SUSY is broken and is yet unknown. Be it as it
may, one has some general hints for the nature of such final
states:
\begin{itemize}
\item Very high mass: since SUSY particles must be heavy
(Tevatron, LEP); \item Large missing energy: at least in all RPC
models due to the existence of a neutral practically
non-interacting LSP.
\end{itemize}
An attempt to construct a search procedure in the most general way
possible, based on these hints, is described in this note.

\subsection{The LSL algorithm}
The K-neighborhood algorithm \cite{hastie} was modified in such a
way that it can cope with the task of finding small deviations
from the simulated expectations, which might result from the
presence of an unspecified signal. In the modified algorithm - the
LSL (Local Spherical Likelihood) \cite{prosso} - each event is
described by N parameters and is represented by a point in a
corresponding N-dimensional space, where the N axes correspond to
the N parameters. The generic name for such a space is the
'\emph{event-space}'. The choice of parameters (i.e. axes) is
crucial as it determines to which type of signal the analysis will
be sensitive. This is the place where model dependence is
introduced into the procedure. Once the parameters (i.e. the axes)
have been chosen one normalizes them (usually the parameters are
mapped in such a way that they are centered at zero and
distributed between zero and one) in order to remove the effect of
variable scaling.

Next, one runs a simulation of all the relevant SM processes and
places each of the simulated events in an event-space, which is
named the '\emph{reference}' space. One proceeds then by
constructing a similar event-space using all data events, this
event-space is named the '\emph{data}' space.

The essence of the algorithm is to look for local accumulations of
events in the '\emph{data}' space, which are absent in the
'\emph{reference}' space. \footnote{Another algorithm which is
designed for the same task is the Sleuth one which has been
developed and used at the Tevatron \cite{sleuth}. The present
algorithm is conceptually simpler.}  In the LSL, in order to
expose the existence of such a local high-density region each of
the data events is placed inside the '\emph{reference}' space and
an N-dimensional sphere is traced around it. The radius of this
sphere is adjusted in such a way that it is the minimum that is
required to contain exactly $N_B$ reference events, where $N_B$ is
a predefined number. The radius of this sphere is then recorded.
Next a sphere with the same radius is traced around the same data
event but this time this is done in the '\emph{data}' space and
the number of data events that are
contained in the sphere $N_D$ is determined.\\
In the absence of signal one expects $N_D \approx N_B$. If signal
were present one would expect $N_D > N_B$ .\\
In order to discern the presence of signal the quantity:
$$\rho(N_B)={N_D-N_B \over \sqrt N_B} $$
is computed. A large value of $\rho$, which is the parameter that
quantifies the local deviation of the density of the data from the
density of the background, is therefore, an indication
for a possible existence of a signal.\\

The numerical value above which $\rho$ can be considered large
enough to constitute an evidence for the presence of a signal in
the data is not well defined at this stage. In order to estimate
this value one makes use of additional SM simulated events (which
are not used for the construction of the \emph{'reference'} space)
and construct a '\emph{null}' space, namely, data-like event-space
in which instead of data one places SM simulated events (without a
signal). One can then repeat the procedure outlined above for the
\emph{'null'} space and get the $\rho$ distribution for the
no-signal hypothesis. The actual value of $\rho$ as computed from
the data, can now be compared with the null-hypothesis and
acquire a meaningful statistical interpretation.\\

\begin{figure}
  \includegraphics[width=16cm]{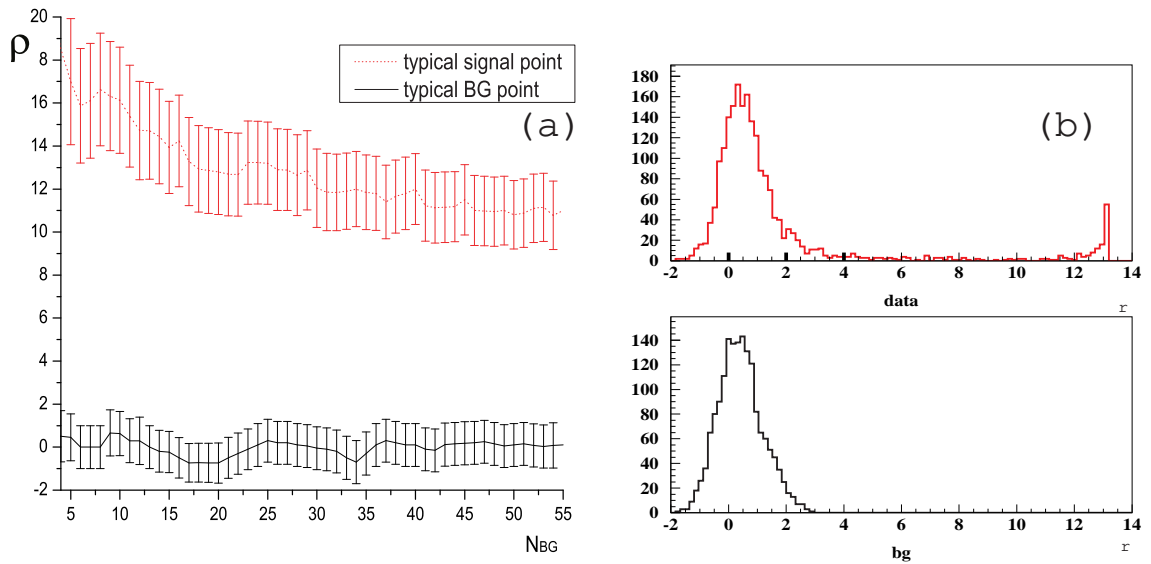}\\
  \caption{a) The variation of $\rho$ as a function of the size of
  the sphere ($N_B$) for a typical background (black line,
  lower band) and signal (red line, upper band) events.
  b)  $\rho^{max}$ distribution for the signal case
  (upper red solid histogram) and for a simulated background,
  (lower black histogram). Note the long tail of high $\rho$ events
  in the signal plot. The small peak at large $\rho^{max}$
  in the upper plot is discussed in the text. }
  \label{first}
\end{figure}

Figure \ref{first}b shows the $\rho$ (for fixed $N_B$ = 21)
distribution for the signal case (upper red histogram) and for a
background case (lower blue histogram). The peak at $\rho \approx
13$ in the signal case is an artifact of the situation in which
the n-dimensional sphere is located at the center of a well
separated cluster of signal events. The sphere contains all the
signal in the cluster and the radius is then artificially enlarged
to include the required 21 background events. Thus, the spheres
around different data points inside this cluster contain the same
set of background and, consequently, signal events. As a result
the $\rho$ value of all the events in this cluster is roughly the
same.

The size of the sphere, namely the numerical value of $N_B$
depends on the number of simulated events as well as on the shape
that the signal cloud takes inside the event space. Since the
second factor is unknown the procedure is repeated for values of
$N_B$ ranging from some minimal value $N_0$ (21 in the present
study) to some fixed value, say 5\% of the number of background
events in the reference space. The maximal attainable $\rho$ in
the series of $\rho(N_B)$, is denoted by $\rho^{max}$.\\
As mentioned above a large value of $\rho^{max}$ is a strong
indication for the existence of a signal in the data.\\
The variation of $\rho$ as a function of $N_B$ is shown in Figure
~\ref{first}a for typical background and signal event. Since
$\rho(N_B)$ is strongly correlated with $\rho(N_B-1)$ the maximum
value of $\rho$ as obtained by this procedure is fairly stable. It
is disadvantageous to evaluate $\rho$ at low values of $N_B$ since
for such values the statistical error is large. It is equally
disadvantageous to evaluate $\rho$ at high values of $N_B$ since
then the radius of the sphere is large and one looses the locality
nature of the analysis.\\

\begin{figure}
  \includegraphics[width=14cm]{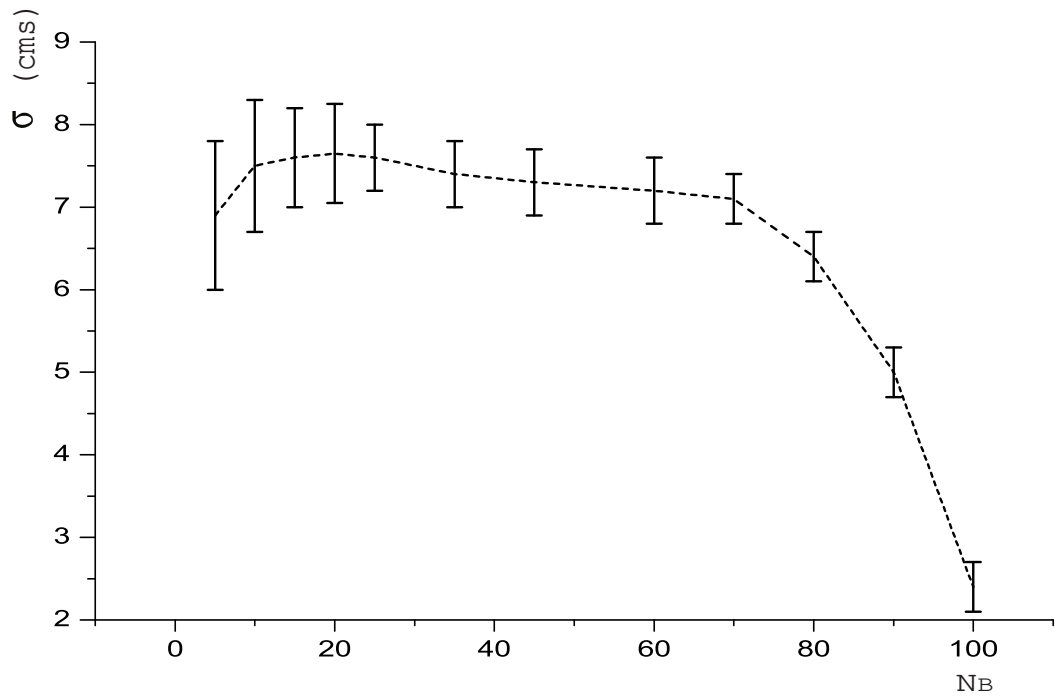}\\
  \caption{The statistical significance of the MSSM search
  analysis as a function of $N_B$.}\label{nb}
\end{figure}

At that point one can select a fixed $N_B$ for which the
attainable $\rho$ are large or continue with $\rho_{max}$ at the
cost of having a variable $N_B$. The results which are presented
below have been obtained using the best $\rho_{max}(N_B)$
attainable provided $N_B>20$. The dependence of sensitivity of the
analysis on $N_B$ is shown in Figure ~\ref{nb} for the case of
GMSB with $\Lambda=170~TeV$, $M=1000~TeV$, $tan \beta=15$ and a
positive $\mu$. One sees, that at small $N_B$ the performance is
not very good because of the statistical fluctuations, while for
large $N_B$ it decreases because of the limited number of signal
events. The optimal $N_B$ in this case is found to be at about 60.

The whole sequence of steps is summarized in the following list:
\begin{enumerate}
\item Choose the parameters (motivated by physics considerations);
\item Apply a set of soft preliminary cuts (to remove irrelevant
events); \item Scale and normalize the events' parameters; \item
Form a '\emph{reference}' space from all relevant SM background
processes; \item Form some '\emph{null}' space by simulating
additional SM simulated events; \item Apply the procedure that was
described before, for obtaining the $\rho$ distributions to the
'\emph{reference}' space and the '\emph{null}' space and obtain
the distribution for $\rho^{max}_{null}$. The number of events in
the '\emph{null}' space should be as large as possible
\footnote{In order to speed up the calculation, the \emph{null}
space was split to several smaller subspaces that are equal in
size to the \emph{data} space. The LSL algorithm is then applied
to each of these subspaces separately, and the average
$\rho^{max}$ is used.}; \item Form the '\emph{data}' space using
preselected data events; \item Apply the procedure that was
described before, for obtaining the $\rho$ distributions to the
'\emph{reference}' space and '\emph{data}' space and obtain the
distribution of the data $\rho^{max}_{data}$;\item Compute
$\sigma(\rho)= {N_{data}-N_{null} \over
\sqrt{N_{null}}}|_{\rho>cut}$ ; where N stands for the number of
events with $\rho>cut$ ~\footnote{The simplest possible
statistical approach is taken here for simplicity sake. Obviously,
if the numbers involved are small a poisson distribution would be
more adequate}, and maximize this value by changing the value of
$\rho_{cut}$.
\end{enumerate}

\section{Implementation}
While the algorithm which was described above can be used to
search for any conceivable signal we study its performance here by
applying it on various RPC SUSY simulated signals. This decision
determines, as was discussed, the selection of parameters  by
which each event is described. On one hand one would like to have
all the relevant parameters that one can think of, but on the
other hand a large number of parameters will necessitates a huge
number of simulated events and will make the procedure either slow
or useless. Hence, only 4 input parameters, with the highest
'separation' power, were selected. In order to conform with
existing analyzes two sets are used: one in which no requirement
on the presence of leptons in the event is set; and another one in
which one lepton is required and its properties are included in
the input parameters. The parameters for the 'no-lepton' case
were:
\begin{itemize}
\item $E_t^{miss}$ - Where $E_t^{miss}$ is the missing
transverse energy of the event;\\
\item $P_t^{jet_1}$ - Where $P_t^{jet_1}$ is the transverse
momentum of the most energetic (transverse direction) jet;\\
\item $P_t^{jet_2}$ - Where $P_t^{jet_2}$ is the transverse
momentum of the second most energetic (transverse direction) jet;\\
\item $\Sigma E_t$ - Where $\Sigma E_t$ is total transverse energy
of the event.
\end{itemize}

In the case of $1-lepton$ channel the 4 input variables are:

\begin{itemize}
\item $E_t^{miss}$ - Where $E_t^{miss}$ is the missing
transverse energy of the event;\\
\item $P_t^{jet_1}$ - Where $P_t^{jet_1}$ is the transverse
momentum of the most energetic (transverse direction) jet;\\
\item $M_{t,l-miss}$ - Where $M_{t,l-miss}$ is the transverse
mass of the lepton-missing momentum system;\\
\item $\Sigma E_t$ - Where $\Sigma E_t$ is total transverse energy
of the event.
\end{itemize}

The SM processes that have been simulated (using Pythia) for this
study consist of the processes: $pp \rightarrow WX$; $pp
\rightarrow ZX$; $pp \rightarrow t \bar t$; $pp \rightarrow
two~jets$. The Equivalent luminosity was set to 10 $fb^{-1}$ and
in order to keep the number of events reasonable, a $p_t$ cut of
200 GeV (via \emph{ckin(3)} \cite{Pythia}) was applied. The effect
of this cut was checked later and verified to be of negligible
importance.

The signal was simulated using Pythia \cite{Pythia}~(for MSSM) and
ISAJET \cite{Isajet} (for GMSB and AMSB). The detector response
was simulated using a fast simulation program \footnote{The
Fortran version of the ATLAS fast simulation program (ATLFAST)
version 2.53}. The ATLAS TDR \cite{TDR} as well as some additional
points were used in this study.

In order to reduce the number of background events in the various
event spaces, a set of preliminary cuts was applied.
\begin{itemize}
\item $E_t^{miss}>~500~GeV$: which is due to the presence of two
LSP in each event;  \item $P_t^{jet_1}>200~GeV$: this cut and the
two that follow reflect the high mass of the expected SUSY
particles; \item $P_t^{jet_2}>100~GeV$; \item $\Sigma
E_t>1500~GeV$;\item $N_{jet}>3$: this cut and the one that follows
are based on the fact that SUSY events are expected to give rise
to long cascade decay chains ; \item $C>0.1$, Where $C$ is the
Circularity of the event.
\end{itemize}

for $1-lepton$ analysis the presence of a lepton with $p_t>10~GeV$
and $|\eta|<2.5$ allows softening some of the cuts. The
preliminary cuts were therefore:
\begin{itemize}
\item $E_t^{miss}>~200~GeV$; \item $N_{jet}>3$; \item
$P_t^{jet_1}>100~GeV$; \item $P_t^{jet_2}>50~GeV$; \item $\Sigma
E_t>200~GeV$; \item $M_{t,l-miss}>80~GeV$: this cut removes most
of the $W+jet$ background.
\end{itemize}

Since the main goal of the present study is the investigation of
the performance of the LSL algorithm no attempt to look for
optimal preselection cuts was done. Rather, the quantities that
were used in \cite{tovey} are used.

\section{Results}
\begin{figure}
  \includegraphics[width=14cm]{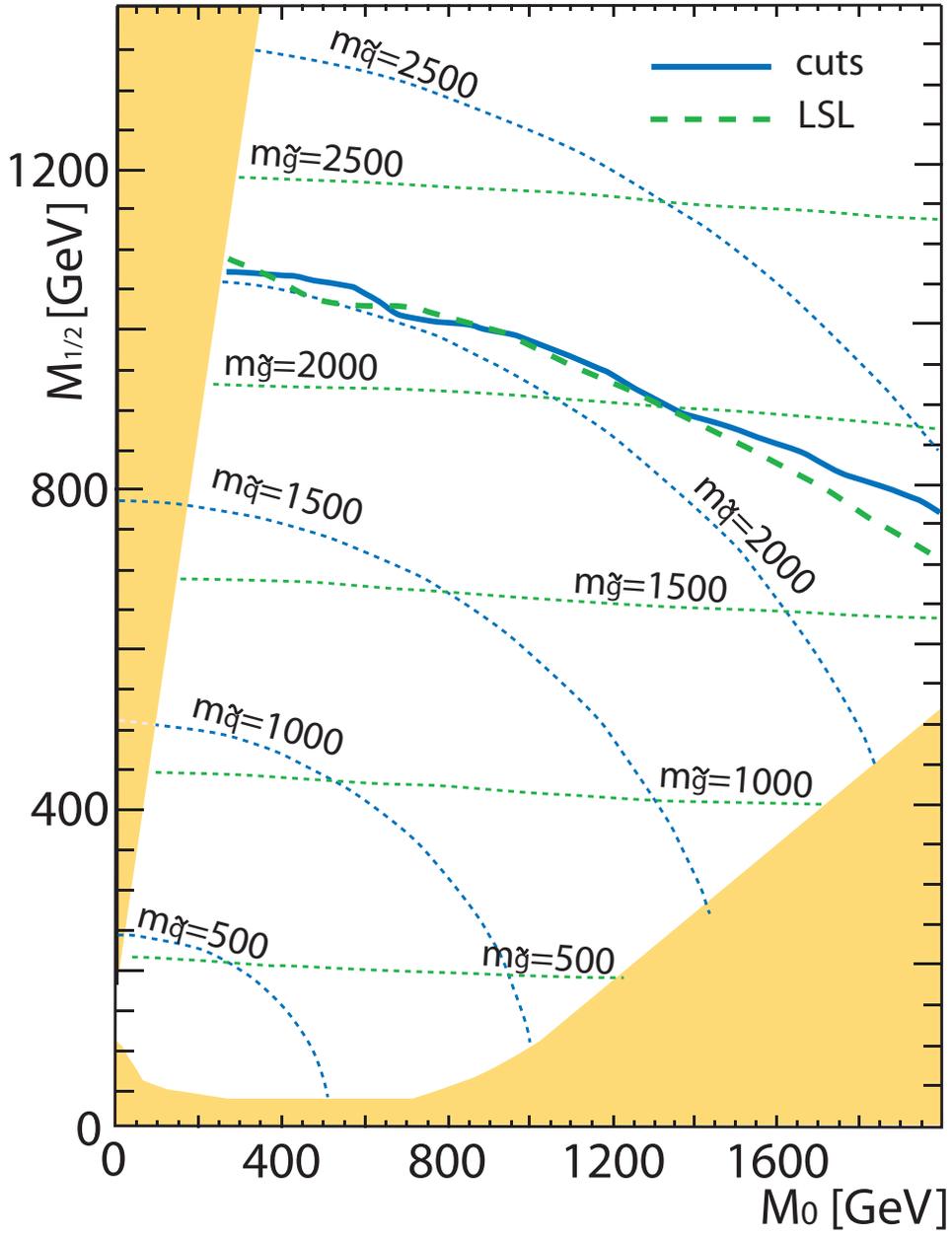}\\
  \caption{The sensitivity reach of ATLAS to MSSM
  signal in the missing energy channel with no requirements set on
  the number of leptons in the event. The solid (black) line is from
  ~\cite{tovey} and the dashed (red) one is the result of the LSL
  algorithm.}\label{meall}
\end{figure}

The sensitivity of ATLAS to predicted signals of several RPC SUSY
models was estimated using the LSL algorithm. In the case of MSSM
and AMSB, it was possible to compare the LSL sensitivity to
conventional procedure. Recently, a comprehensive evaluation of
ATLAS's sensitivity to a MSSM signal was performed \cite{tovey}.
On top of introducing a new channel, namely, the missing energy
channel with no requirement on leptons, which proved to be the
best search channel, this study also introduced a sophisticated
automatic cut optimization procedure which is based on the Simplex
algorithm. Figure ~\ref{meall} is a comparison between this
technique in which the signal is simulated and cuts are optimized
in numerous points and the LSL algorithm in which no simulation of
the signal was used at all.

\begin{figure}
  \includegraphics[width=14cm]{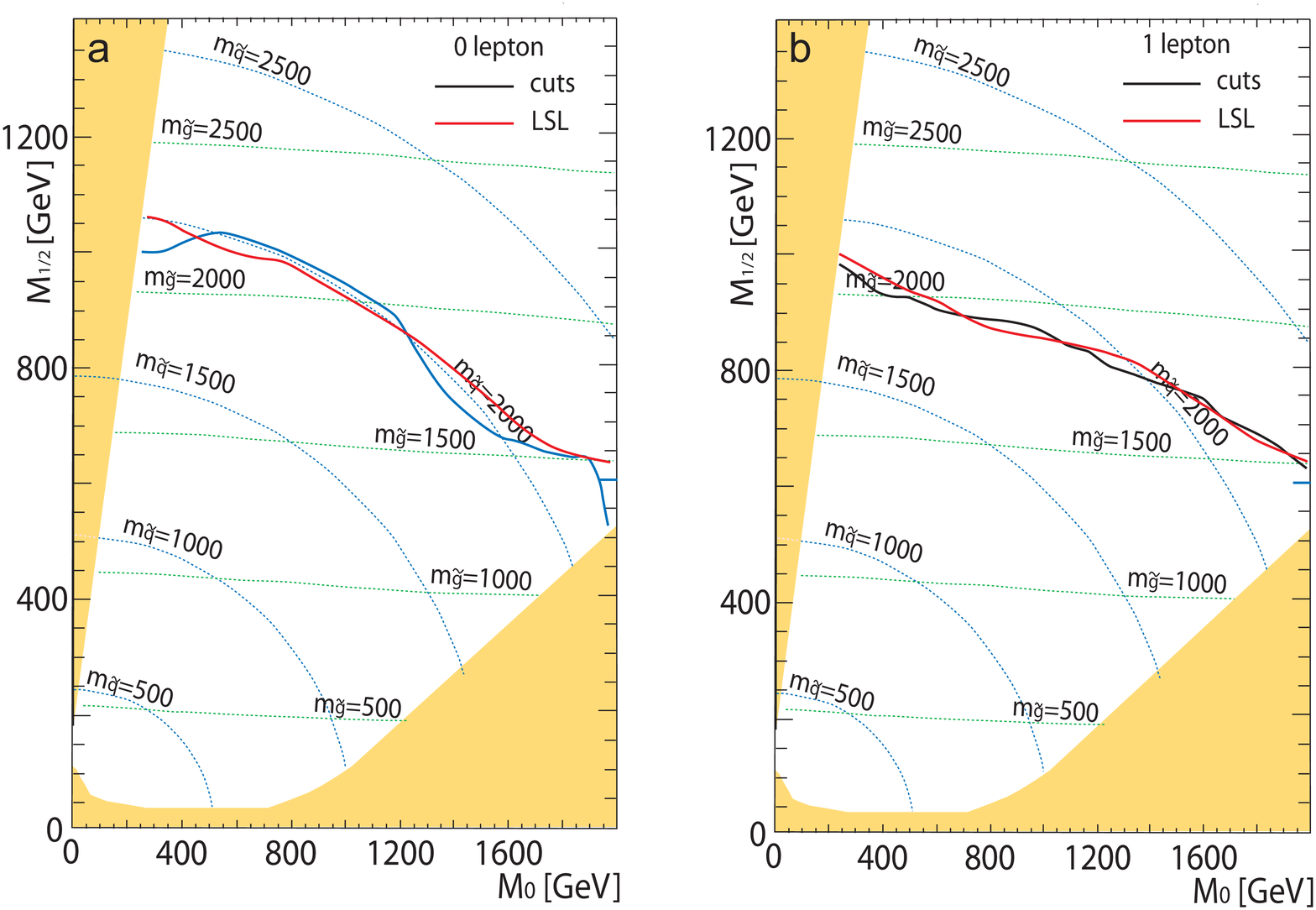}\\
  \caption{The sensitivity reach of ATLAS to MSSM
  signal in the missing energy channel with no leptons in the
  event (a) and with one leptons in the event (b).
  The solid (black) line is from ~\cite{tovey} and the dashed
  (red) one is the result of the LSL
  algorithm.}\label{mssmlt}
\end{figure}

Figure ~\ref{mssmlt}a is a similar comparison between the two
methods when only events with no leptons are considered. A
somewhat complementary case, namely the case when events are
required to have one isolated energetic lepton is shown in Figure
~\ref{mssmlt}b. An attempt to combine these two searches was also
carried out. Such a combination, which is similar to the one
applied in Higgs boson searches at LEP, is expected to lead to an
improved sensitivity. However, the improvement which was obtained
was only marginal.

\begin{figure}
  \includegraphics[width=14cm]{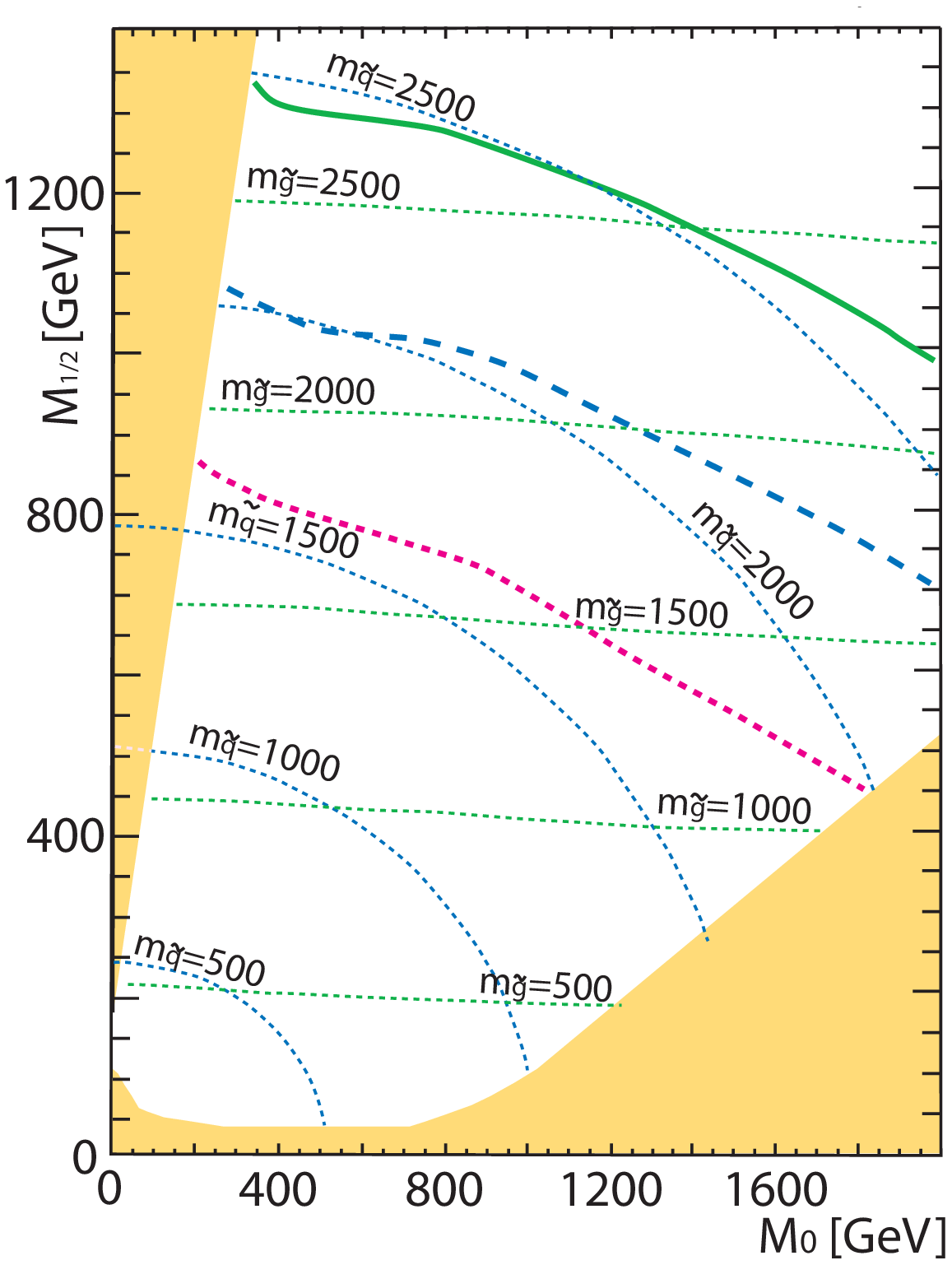}\\
  \caption{The sensitivity reach of ATLAS to MSSM
  signal as estimated using the LSL
  algorithm for luminosities of 1, 10 and 100 $fb^{-1}$ .}
  \label{finalmssm}
\end{figure}

Generally speaking one may conclude from these plots that the
sensitivity of the two methods is comparable. Yet one should bare
in mind that {\bf the LSL algorithm did not make any use of
simulated signal}. For completeness sake the sensitivity of ATLAS
for a MSSM signal as estimated with the LSL algorithm for
luminosities of 1, 10 and 100 $fb^{-1}$ is presented in
Figure \ref{finalmssm}.\\

\begin{figure}
  \includegraphics[width=14cm]{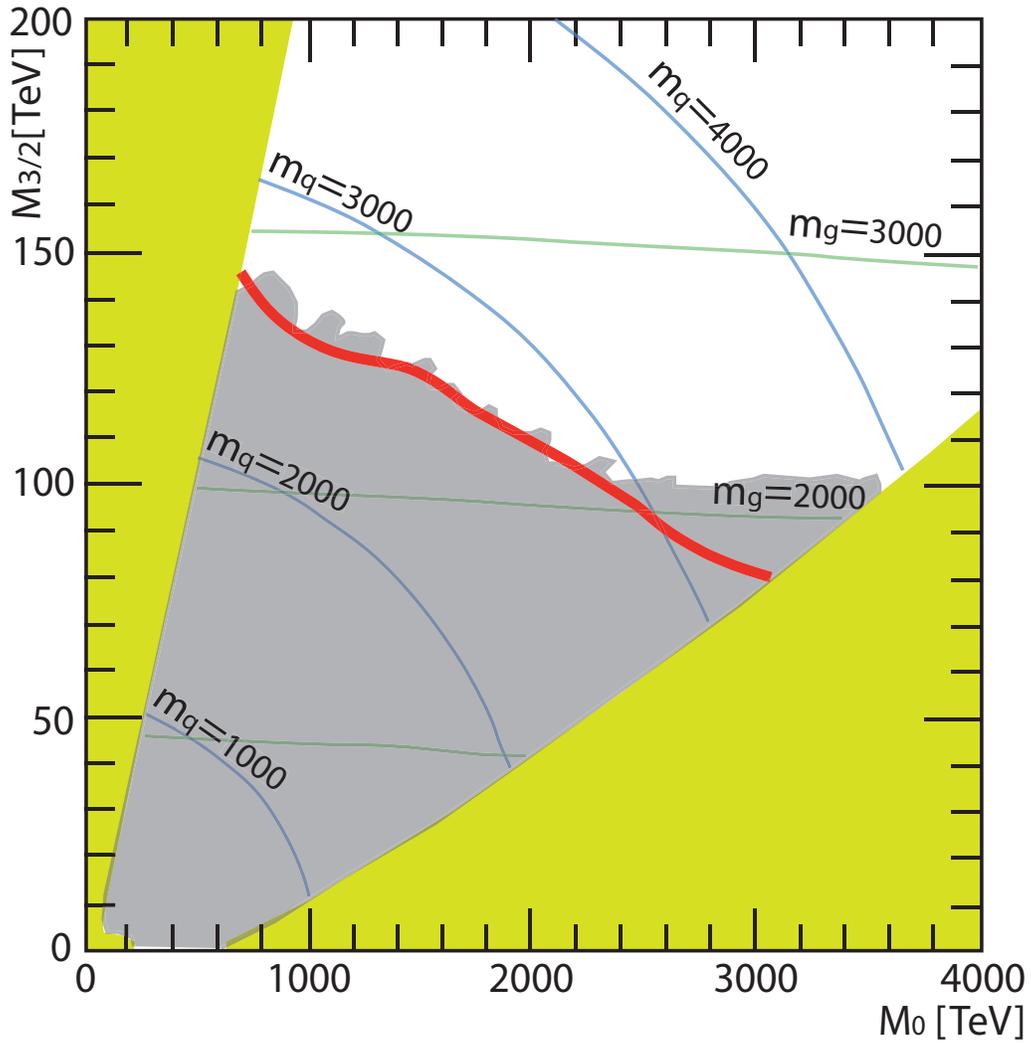}\\
  \caption{A comparison between the LSL sensitivity and the published
  results \cite{amsb} of the search for AMSB signal.
  The blue circled area
  represents the estimated sensitivity of the dedicated search while
  the thick dotted black line is the LSL sensitivity limit.}
  \label{amsbcomp}
\end{figure}

A Study of ATLAS sensitivity to possible AMSB signal was carried
out by Barr, Allanach, Lester and Parker \cite{amsb}. In order to
extract a signal a set of quantities were selected and were
subjected to various cuts. 10 sets of such cuts were used for the
various analyzes that have been done: 0-lepton, 1-lepton,
2-oppositely charged lepton etc.. In order to compare the LSL
performance with this analysis while keeping the wide-scope
approach, the \emph{null} and \emph{reference} spaces that were
used in the MSSM case were used also here. No modification
whatsoever was introduced except for the introduction of a
simulated AMSB signal into the \emph{data} space instead of the
MSSM one. The comparison of the $E_{miss}$ analyzes is shown in
figure \ref{amsbcomp}.

The two analyzes are again comparable except for the right side of
Figure \ref{amsbcomp} where the LSL performance is inferior to the
conventional technique. This behavior is related to the number of
events with large number of jets and the differences in their
simulation between Herwig (used by \cite{amsb}) and Pythia
(background simulation in LSL case). Note that the sensitivity
region here is
estimated by  ${S \over \sqrt{B}}>5$ and $S>10$. \\

For completeness a three-luminosity contour, with 1, 10 and 100
$fb^{-1}$ is also given when the sensitivity is estimated with a
more stable estimator, namely requiring ${S \over \sqrt{S+B}}>5$
where S and B are the number of signal and background events
respectively.

\begin{figure}
  \includegraphics[width=14cm]{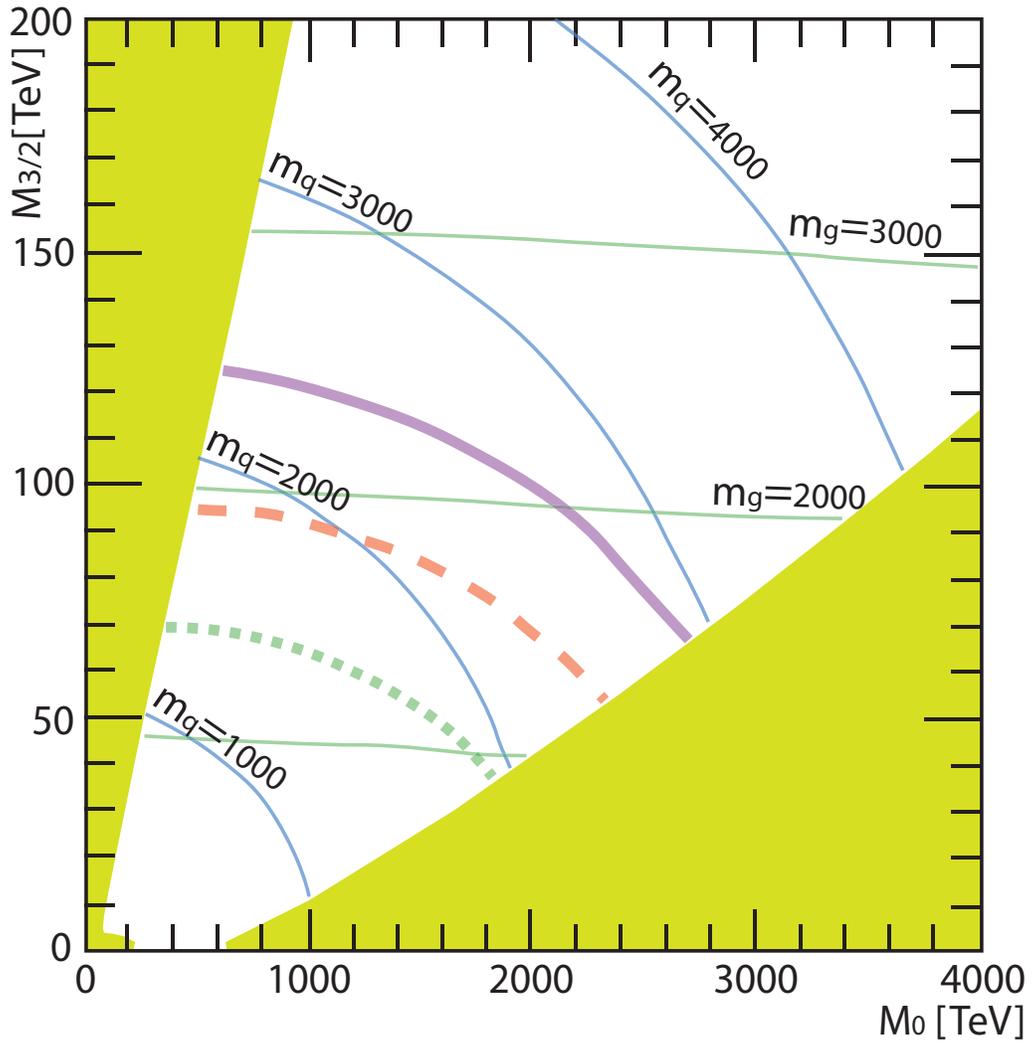}\\
  \caption{The sensitivity reach of ATLAS in the AMSB parameter space
  for luminosities of 1, 10 and 100 $fb^{-1}$.}
  \label{amsb0l}
\end{figure}

\begin{figure}
  \includegraphics[width=14cm]{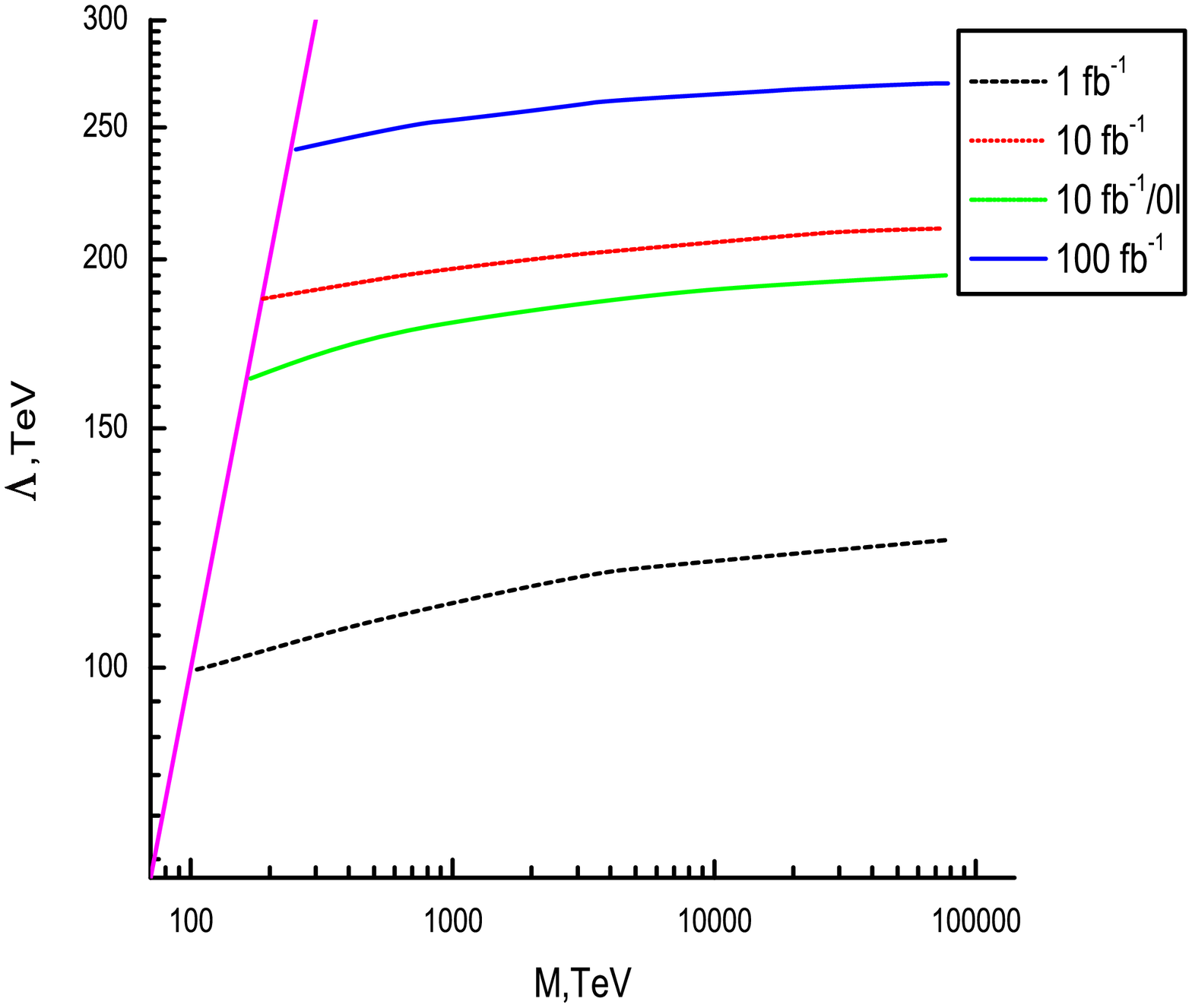}\\
  \caption{Sensitivity reach of ATLAS for GMSB signal.}
  \label{gmsb}
\end{figure}

A similar procedure was repeated for the GMSB case. The LSL inputs
were left unchanged and the estimated ATLAS sensitivity is shown
in Figure ~\ref{gmsb}. It is found again to be comparable to the
one
which was obtain with a naive set of conventional cuts ~\cite{lund}.\\

The LSL is basically looking for deviations of the data from the
SM expectation, as predicted by the simulation. As such it might
be sensitive to the quality of the simulation. Defects in the
simulation can easily be misinterpreted as indication of a signal.
Some preliminary studies of the stability of the algorithm under
artificial distortion of the simulation are described in Appendix
A.

\section{Conclusion}
The LSL sensitivity was shown to be comparable to the one
attainable by carefully adjusting the cuts to a signal of
well-known characteristics. It is possible that a more
sophisticated analysis, which is based on likelihood or artificial
neural networks, will be superior to the LSL algorithm. Yet one
should bare in mind that {\bf the LSL algorithm did not make any
use of simulated signal}. Hence, the LSL will be able to observe
signals of unpredicted nature and once such deviation are exposed;
they will be studied using all available analysis tools.

\section{Acknowledgments}
We would like to thank the members of the Weizmann group: Eilam
Gross, Arie Melamed-Katz, Michael Riveline and Lidija Zivkovic for
their help in discussing the various aspect of the present work.
We would also like to thank Dan Tovey for his help in making this
paper clearer. We are obliged to the Benoziyo center for High
Energy Physics for their support of this work. This work was also
supported by the Israeli Science Foundation(ISF) and by the
Federal Ministry of Education, Science, Research and Technology
(BMBF) within the framework of the German-Israeli Project
Cooperation in Future-Oriented Topics(DIP).

\begin{figure}
  \includegraphics[width=16cm]{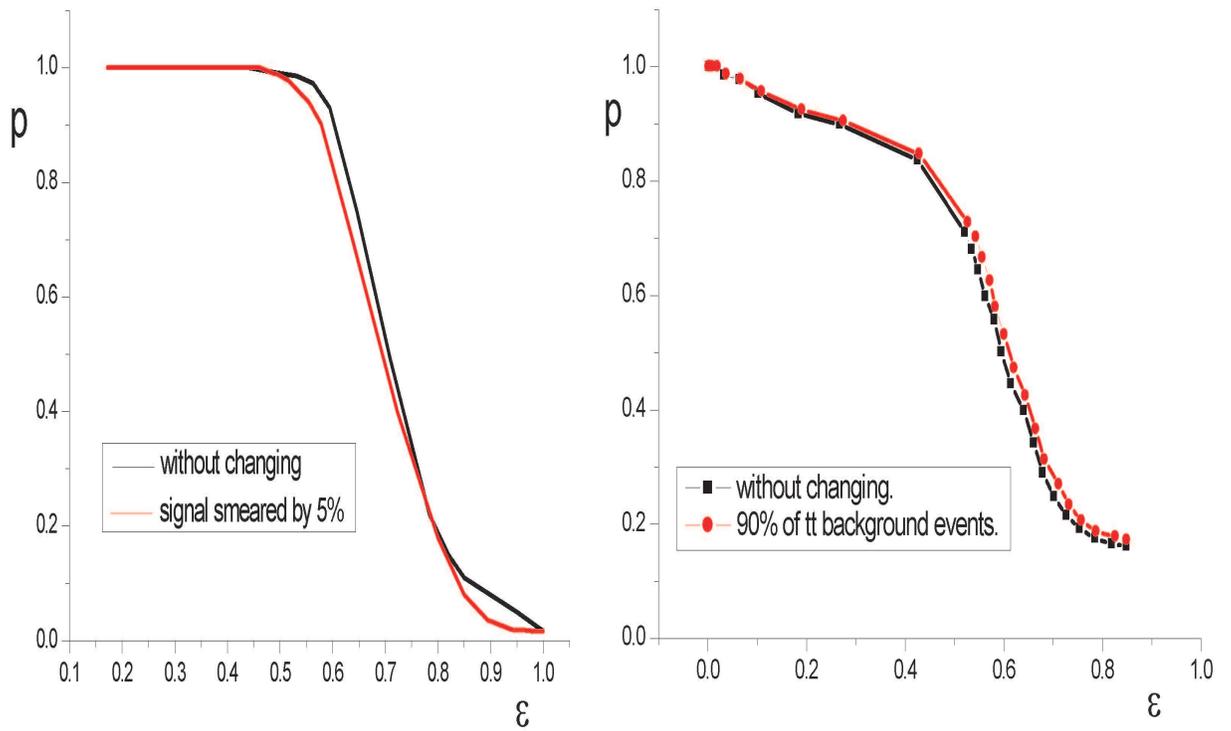}\\
  \caption{The efficiency vs. purity curve under normal conditions
  and after scaling down the jet energy in the 'data' by 5\%.
  }\label{systematics}
\end{figure}

\vskip 1cm
\appendix{\Large \bf Appendix A}\\
\vskip 0.5 cm Differences between the data and the simulated
signals trigger the LSL algorithm. Such differences may indicate
the presence of a signal but might result also from bad modelling
of the detector and/or from bad modelling of the various SM
processes. In order to evaluate the effect of the later sources
few preliminary studies have been done. The first test checked the
sensitivity to energy calibration. The  energy of the 'measured'
events (i.e. those in the \emph{data} space) was scaled down by
5\% while that of the simulated SM (the \emph{reference} space)
was left untouched. The efficiency/purity of the signal selection
procedure for a MSSM signal under these conditions was compared to
the one which was obtained under normal conditions. The results
are shown in Figure \ref{systematics}a\\

Another potential source of fake signal is mismodelling of SM
processes. In order to evaluate the importance of this source of
trouble the $t \bar t$ process was scaled down by 10\% in the
\emph{reference} space, leaving the 'data' richer in $t \bar t$ by
10\% more that 'predicted. The efficiency vs purity performance
curve of the LSL is shown in Figure \ref{systematics}b.\\

One may conclude from these two tests that the algorithm is fairly
stable to the tested forms of distortion.


\begin{thebibliography}{7}
\bibitem{hastie} T. Hastie and R. Tibshirani, IEEE PAMI, 18
607-616, 1996.
\bibitem{prosso} E. Prosso, MSc. Thesis, Weizmann Istitute of
Science, July 2002.
\bibitem{sleuth} B. Knuteson HEP-EX/0105027,
0006011.
\bibitem{Pythia}
Pythia 6.2 Physics and Manual.Torbjorn Sjostrand, Leif Lonnblad,
Stefen Mrenna, Peter Scands LU-TP 01-21 (2002).
\bibitem{Isajet}
ISAJET 7.63 A Monte-Carlo Event Generator, Frank E. Paige and
Serban D. Protopescu.
\bibitem{TDR}
ATLAS Detector and Physics Performans TDR, CERN LHCC/99-14 vol II.
\bibitem{knn} T. Hastie and R. Tibshirani, IEEE PAMI, 18, 607-616,
1996.
\bibitem{tovey} D.R. Tovey, 'Inclusive SUSY Searches and Measurements at ATLAS',
 Eur. Phys. J. C{\bf4}(2002) N4, SN-ATLAS-2002-020.

\bibitem{amsb} Discovering anomaly-mediated supersymmetry at the
LHC\\
Barr, A J; Allanach, Benjamin C; Lester, C G; Parker, M A;
Richardson, Peter. \\
hep-ph/0210182, J. High Energy Phys. 03 (2003) 045
\bibitem{lund} E. Duchovni, $3^{ed}$ ATLAS Physics Workshop, Lund,
September 2001.

\end{thebibliography}
\end{document}